\documentclass[amsmath,amssymb,12pt]{revtex4}
\usepackage{graphicx}
\usepackage{pstricks}
\def\lsim{\mathrel{\rlap{\lower4pt\hbox{\hskip1pt$\sim$}}
    \raise1pt\hbox{$<$}}}
\def\gsim{\mathrel{\rlap{\lower4pt\hbox{\hskip1pt$\sim$}}
    \raise1pt\hbox{$>$}}}
\def\sqr#1#2{{\vcenter{\vbox{\hrule height.#2pt
         \hbox{\vrule width.#2pt height#1pt \kern#1pt
         \vrule width.#2pt}
         \hrule height.#2pt}}}}
\newcommand{\beq}{\begin{equation}}
\newcommand{\eeq}{\end{equation}}
\newcommand{\bea}{\begin{eqnarray}}
\newcommand{\eea}{\end{eqnarray}}

\def\klm#1#2#3{k^{(#1)}_{(#2)#3}}

\def\figdir#1{figs/#1}
\def\etal{{\it et al.}}

\begin{document}

\title{\normalsize Analysis of Cosmic Microwave Background Radiation
  in the Presence of Lorentz Violation\footnote{%
    Presented at the 2006 Wisconsin Space Conference,
    Milwaukee, Wisconsin, August 10-11, 2006.}}
\author{Matthew Mewes}
\affiliation{
  Physics Department, Marquette University,
  Milwaukee, WI 53201, U.S.A.}

\begin{abstract}
  We examine the effects Lorentz violation
  on observations of cosmic microwave 
  background radiation.
  In particular,
  we focus on changes in polarization
  caused by vacuum birefringence.
  We place stringent constraints
  on previously untested violations.
\end{abstract}

\maketitle

Today, Einstein's special relativity
is understood to be a consequence of
the more fundamental principle of
Lorentz invariance.
All modern physical theories
are explicitly Lorentz invariant.
The prominent position
of Lorentz symmetry
at the foundations of known
physics makes its verification essential.
Furthermore,
violations of Lorentz symmetry
have been identified as
a promising avenue for searches for
new physics arising from a 
unified description of nature,
such as strings
and other attempts at
quantized gravity  
\cite{kost,ck,qg}.
Such violations are expected be
exceedingly tiny,
but a number of experiments have
achieved the necessary sensitivity
to probe relativity at interesting levels 
\cite{cpt04}.

Among the most sensitive experiments are
those that involve polarized light
from sources at cosmological distances 
\cite{cfj,km1,km2,km3,grb,dispersion,feng}.
The Cosmic Microwave Background (CMB)
marks the limit of the observable universe
\cite{cmbrev},
and therefore provides an excellent
opportunity for tests of Lorentz invariance.
Here we consider the effects of
relativity violations on the CMB
and use it to place constraints on a 
wide class of unconventional effects.

A general theoretical framework
known as the Standard-Model Extension (SME)
has been developed to 
facilitate the study of
Lorentz-symmetry violations
\cite{ck}.
The SME is constructed out of the
usual Standard Model of particle physics,
augmented by all reasonable additions
that are consistent with 
experimental constraints.
The SME provides the low-energy limit of
any realistic theory and
characterizes Lorentz violation
independent of its origin.
The SME provides the theoretical
backing for tests involving 
atomic clocks \cite{ccexpt},
neutral mesons \cite{hadrons},
spin-polarized materials \cite{spinpol},
particle traps \cite{penning},
muons \cite{muons},
neutrinos \cite{neutrinos},
and
photons.
Photon tests include 
modern versions of the
classic Michelson-Morley experiment \cite{cavities}
and searches for changes in
polarization of light from
distant sources \cite{km1,km2,km3}.
The latter tests yield sensitivities
comparable to the best tests in
any sector of the SME.

Changes in polarization arise
out of an effect known as
``cosmic birefringence.''
This occurs when certain forms
of relativity violations cause
light to propagate as the superposition
of two waves that differ in velocity
and polarization.
The difference in velocity causes
the net polarization of the composite
wave to evolve as the light propagates in vacua.
Searches for this change in light
from sources at cosmological distances
yield extreme sensitivity to
violations of Lorentz symmetry
due to the long propagation times
over which the tiny effects can accumulate.
CMB radiation was created around 
13 billion years ago,
and has propagated more or less unhindered since.
It constitutes the oldest unscattered
light available to observation
and is therefore an ideal source
for birefringence searches.

CMB radiation is released during the
epoch of recombination,
a period when the Universe has cooled
to a temperature at which atoms can form
\cite{cmbrev}.
At this time,
the Universe suddenly became transparent,
and photons were now free to travel great
distances without scattering.
The CMB thus provides a snapshot
of the Universe during a early stage
of its evolution, approximately
300,000 years after the big bang.
Signatures of the conditions 
in the early Universe are imprinted 
in the photons at the surface of last scatter,
and provide a powerful probe into early cosmology.

While the Universe was extremely
homogeneous during recombination,
tiny variations in the temperature
of the CMB blackbody spectrum
have been firmly established
by a host of experiments
\cite{cobe,boomtemp,wmaptemp}.
These variations point to small
inhomogeneities in temperature
and density of the primordial plasma.
These inhomogeneities
give rise to the possibility of
polarized scattering
\cite{cmb-pol}.
In a homogeneous universe,
there are no preferred directions.
Consequently, scattering always
results in zero total polarization.
The argument fails if there exists 
a gradient in the density of
scattering particles.
This results in preferential scattering
directions that depend on the polarization 
of the incident photons.
The effect is that light
scattering in certain directions
with respect to the inhomogeneities
can have a net polarization.
Furthermore,
these should correlate in a predictable
way with the variations in temperature.

The CMB is generally decomposed into
temperature $T$
and two types of polarization,
the so-called $E$ and $B$ types.
The $E$ and $B$ modes characterize
any pattern of linear polarization
coming from all points on the sky.
In general, we may also have
circular polarization, type $V$,
but this is not produced 
during recombination according to 
conventional physics
\cite{cmb-pol}.
The breakdown into $E$ and $B$
is convenient since only the $E$ type
is expected to be correlated with
the variations in temperature.
Numerous observations
\cite{boompol,dasi,cbi,capmap,wmappol}
have confirmed the existence of $E$
polarization in the CMB and
a correlation with $T$.

Next we examine the effects of
birefringence on polarization in the CMB.
Working within the SME framework,
we construct all possible Lorentz-violating
additions to conventional electrodynamics
that can cause birefringence.
A more detailed discussion is provided in Ref.\ \cite{km3}.
Here we summarize the basic results.
The Lorentz-violating 
modifications to electrodynamics can 
be classified in a similar manor as
CMB polarization, as types $E$, $B$, and $V$.
The changes in polarization are governed by
a set of coefficients for Lorentz violation,
which we write as
$\klm{d}{V}{lm}$, $\klm{d}{E}{lm}$, and $\klm{d}{B}{lm}$
for the three types of violations.
These coefficients
completely characterize the effects
that violations of relativity can have
on the polarization of the CMB.
Nonzero values of these coefficients
lead to birefringence, which will
cause the pattern of polarization
on the sky to change during the time
it takes CMB radiation to propagate to Earth.

Some generic effects arise from 
the various coefficients.
For example, roughly speaking,
the index $d=3,4,5,\ldots$ characterizes 
the frequency dependence of the changes
in polarization 
\cite{fn1}.
For all  $d>3$ coefficients,
higher frequency means stronger 
birefringence and a bigger change
in the polarization.
Only the $d=3$ case causes changes
in polarization that are independent of frequency.
There are several signatures of this
frequency dependence that may be
detectable in CMB experiments that
could provide signals of Lorentz violation.
For example,
typically CMB observations are made
over frequency bands, not single frequencies.
If the frequency dependence is large,
the difference in the polarization 
at different frequencies in the band 
may be great enough so that the
measured polarization
(the average over the frequency band)
is effectively reduced due to a loss of
coherence across the band.

The indices $l$ and $m$ arise from a
(spherical-harmonic) multipole
decomposition of the Lorentz violation
\cite{fn2}.
The value of these indices tell us
something about the directional dependence
of the resulting birefringence.
Higher values of $l$ and $m$ say that
the changes to the CMB polarization
are more dependent on the location on the sky
from which the radiation originates.
Only the $l=0$ and $m=0$ coefficients
cause a uniform change in the polarization
across the entire sky. 
This unconventional direction dependence 
also leads to some signals for Lorentz
violation that could be sought 
in CMB experiments,
such as reductions in net 
measured polarization.

One key feature of birefringence is
that it causes mixing of the $E$, $B$,
and $V$ type polarizations.
Several generic signals for Lorentz
violation arise from these mixings.
For example,
the $\klm{d}{V}{lm}$ coefficients 
result in mixing between the
$E$- and $B$-type polarizations.
For the CMB,
this can cause unconventional behavior,
such as a correlation between $T$ and $B$.
Recall that in the conventional case
we expect there to be a connection between
the $T$ and $E$ patterns on the sky
since the mechanism generating them are linked.
If birefringence causes
the $E$ polarization to be converted to
$B$ polarization as it propagates to Earth
then a correlation between $T$ and $B$ might arise.

The $\klm{d}{E}{lm}$ and $\klm{d}{B}{lm}$
coefficients cause a mixing of 
$E$, $B$, and $V$ polarization.
A possible signal of this type of mixing
is the generation of significant
$V$ modes, \it i.e.\rm, circular polarization.
As discussed above,
the physics which produces the
net polarization in the CMB is
expected to produce only linear polarization.
Observations of a large $V$ component in
the CMB may point to birefringence
caused by this type of mixing.

The numerous types of violations and
there resulting effects makes a
comprehensive search for birefringence
in the CMB impractical.
Here we provide a limited systematic
comparison to available data.
In principle,
we could incorporate all available
polarization data,
but the complicated
frequency dependence makes an analysis
of this type difficult.
So, as a first step,
we limit our search to the
BOOMERANG experiment
\cite{boompol},
whose data are particularly
well suited for our purposes.
They report polarization measurements
for types $E$ and $B$ for a single
narrow frequency band at about 145 GHz.
This relatively high frequency
implies a greater sensitivity
to most Lorentz violating effects.
While incorporating lower-frequency data,
will decrease the size of the errors
found below, it is not likely
to significantly effect the
overall sensitivity.

\begin{figure*}
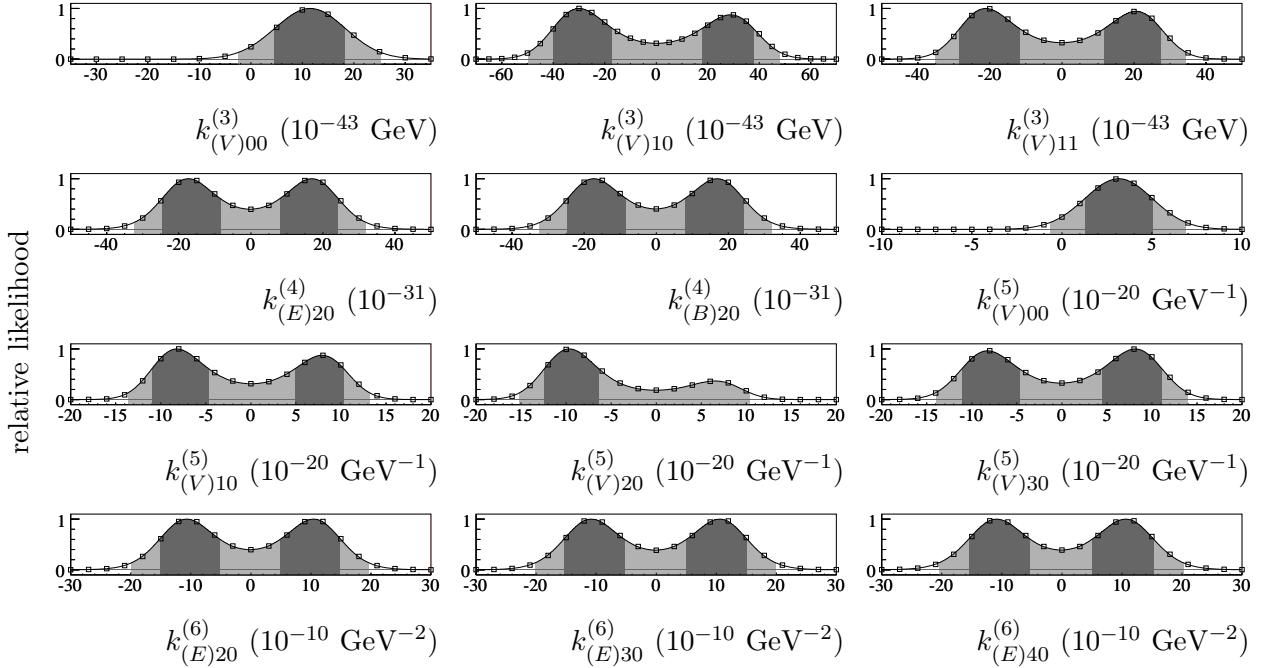

  \rput{90}(0,0){relative likelihood}
  \hfill
  \begin{tabular*}{0.97\textwidth}{@{\extracolsep{\fill}}rrr}
    \includegraphics[width=0.31\textwidth]{\figdir{boom_v300.eps}}&
    \includegraphics[width=0.31\textwidth]{\figdir{boom_v310.eps}}&  
    \includegraphics[width=0.31\textwidth]{\figdir{boom_v311.eps}}\\
    $\klm{3}{V}{00}\ (10^{-43} \mbox{ GeV})$&
    $\klm{3}{V}{10}\ (10^{-43} \mbox{ GeV})$&
    $\klm{3}{V}{11}\ (10^{-43} \mbox{ GeV})$\\[6pt]
    \includegraphics[width=0.31\textwidth]{\figdir{boom_e420.eps}}&
    \includegraphics[width=0.31\textwidth]{\figdir{boom_b420.eps}}&
    \includegraphics[width=0.31\textwidth]{\figdir{boom_v500.eps}}\\   
    $\klm{4}{E}{20}\ (10^{-31})$&
    $\klm{4}{B}{20}\ (10^{-31})$&
    $\klm{5}{V}{00}\ (10^{-20} \mbox{ GeV}^{-1})$\\[6pt]
    \includegraphics[width=0.31\textwidth]{\figdir{boom_v510.eps}}&
    \includegraphics[width=0.31\textwidth]{\figdir{boom_v520.eps}}&
    \includegraphics[width=0.31\textwidth]{\figdir{boom_v530.eps}}\\ 
    $\klm{5}{V}{10}\ (10^{-20} \mbox{ GeV}^{-1})$&
    $\klm{5}{V}{20}\ (10^{-20} \mbox{ GeV}^{-1})$&
    $\klm{5}{V}{30}\ (10^{-20} \mbox{ GeV}^{-1})$\\[6pt]
    \includegraphics[width=0.31\textwidth]{\figdir{boom_e620.eps}}&
    \includegraphics[width=0.31\textwidth]{\figdir{boom_e630.eps}}&
    \includegraphics[width=0.31\textwidth]{\figdir{boom_e640.eps}}\\   
    $\klm{6}{E}{20}\ (10^{-10} \mbox{ GeV}^{-2})$&
    $\klm{6}{E}{30}\ (10^{-10} \mbox{ GeV}^{-2})$&
    $\klm{6}{E}{40}\ (10^{-10} \mbox{ GeV}^{-2})$
  \end{tabular*}
  \caption{\label{plots}
    A sample of relative likelihood 
    versus coefficients for Lorentz violation.
    Boxes indicate calculated points,
    and the curve is the smooth 
    extrapolation of these points.
    The dark-gray indicates the
    estimated 68\% confidence level,
    and the light-gray region shows the 95\% level.}
\end{figure*}

Figure \ref{plots}
shows our comparison of
the BOOMERANG measured polarization 
to what we would expect
in the event of nonzero Lorentz violation.
The shaded regions indicate the ranges
of coefficients which are preferred by
the data at the 68\% and 95\% confidence level.
Here we consider only one 
nonzero coefficient at a time,
although, in principle,
any combination of coefficients may 
exist in nature.

As can be seen in Fig.\ \ref{plots},
we find some general features.
All of the coefficients that we 
tested against BOOMERANG data prefer 
nonzero values at the 68\% level,
but are consistent with zero at 95\%.
This implies that we have a
positive signal for Lorentz violation
at the one-standard-deviation level.
Future tests will be required to
determine the reliability of this signal.
We can also use the 95\% constraints
to place more conservative bounds on
the coefficients for Lorentz violation.
Table \ref{table} summarizes the bounds
we obtain for the lowest values of $d$.

Some results for the $d=3$ and $d=4$ cases exist
\cite{cfj,km1,km2,feng}.
Our positive signal for the $d=3$
coefficients are consistent with
the signal found by Feng \etal
in an analysis utilizing both
BOOMERANG and WMAP
\cite{feng}.
Our result also constitutes
an improvement, by a factor of $\sim 2$,
on previous bounds for $d=3$ coefficients,
which were obtained by from 
observations of distant radio galaxies
\cite{cfj}.
For most the $d=3$ coefficients,
our analysis constitutes the most
sensitive tests to date.

The bounds on $d=4$ coefficients
are comparable to those
obtained from observations of distant galaxies
\cite{km1}.
The highest sensitivities are
achieved by observations of
gamma-ray bursts
\cite{km2}.
While not as sensitive as this search, 
the CMB provides the first test 
utilizing non-point-like sources.
The advantage of a global source like
the CMB is that it can constrain
a much larger region of coefficient space,
providing a more robust limit.

Observations of the CMB have already
provided exceptional tests of
early cosmology and 
the evolution of the Universe.
As demonstrated above,
the CMB is also a powerful probe into
the nature of spacetime itself.
Current data have the ability
to investigate fundamental 
spacetime symmetries with extreme precision.
We not only find that the CMB yields
constraints that are comparable or
surpass any previous test,
but a hint of Lorentz violation
exists in current CMB data.
Forthcoming observations \cite{sport,planck}
will certainly yield improved sensitivity
and are likely to provide
excellent tests of special relativity.

\begin{table}
  \begin{tabular}{c|cccc}
    \hline
    $d$ & 3 & 4 & 5 & 6 \\
    bounds \hfill (GeV$^{4-d}$) & $10^{-42}$ & $10^{-30}$ & $10^{-19}$ & $10^{-9}$  \\
    \hline
  \end{tabular}
  \vskip3pt
  \caption{
    Order of magnitude estimate
    of bounds on coefficients with
    index $d$ from BOOMERANG.
  }
  \label{table}
\end{table}

\vskip20pt
\noindent
\bf Acknowledgements: \rm
This work is supported in part by
the Wisconsin Space Grant Consortium.
The author would like to thank collaborator
V.A.\ Kosteleck\'y.


\begin{thebibliography}{99}
\bibitem{fn1}
  The $V$-type coefficients 
  $\klm{d}{V}{lm}$ have odd values of $d$,
  while $\klm{d}{E}{lm}$, and $\klm{d}{B}{lm}$
  correspond to even values.
\bibitem{fn2}
  The indices $l$ and $m$
  can take on any value obeying the
  restrictions $0 \leq l \leq d-2$ for $V$,
  $2 \leq l \leq d-2$ for $E$ and $B$, 
  and $|m| \leq l$ for all three.
\end{thebibliography}
\end{document}